\documentclass[12pt]{article}
\usepackage[margin=1in]{geometry}

\usepackage{amssymb}
\usepackage{amsmath}
\usepackage{amstext}
\usepackage{graphicx,epsfig}
\usepackage{epsfig}
\usepackage{verbatim} 
\usepackage{fancyhdr}
\usepackage{fancybox}
\usepackage{color}
\usepackage{ulem,bbold}
\usepackage{enumitem}
\usepackage{caption}
\usepackage{subcaption}
\usepackage{bbm}
\usepackage{parskip}
\usepackage{cite}

\newcommand{\Comment}[1]{{}}
\definecolor{MyDarkBlue}{rgb}{0.15,0.15,0.45}
\usepackage[linktocpage=true]{hyperref}
\hypersetup{
colorlinks=true,
citecolor=MyDarkBlue,
linkcolor=MyDarkBlue,
urlcolor=MyDarkBlue,
}

\newcommand{\be}{\begin{equation}}
\newcommand{\ee}{\end{equation}}
\newcommand{\bea}{\begin{eqnarray}}
\newcommand{\eea}{\end{eqnarray}}

\numberwithin{equation}{section}

\begin{document}

\thispagestyle{empty}

\begin{center}
{\Large \bf{On Causality Conditions in de Sitter Spacetime}}
\end{center} 

\vspace{1truecm}
\thispagestyle{empty}
\centerline{\large Noah Bittermann,\footnote{\href{mailto:nb2778@columbia.edu}{\texttt{nb2778@columbia.edu}}} 
Daniel McLoughlin\footnote{\href{mailto:dcm2183@columbia.edu} {\texttt{dcm2183@columbia.edu}}}
and Rachel A. Rosen\footnote{\href{mailto:rar2172@columbia.edu} {\texttt{rar2172@columbia.edu}}}}

\vspace{.5cm}

 \centerline{{\it Center for Theoretical Physics, Department of Physics,}}
 \centerline{{\it Columbia University, New York, NY 10027}}

 \vspace{1cm}
\begin{abstract}
\noindent
 We carefully consider the Shapiro time delay due to black holes and shockwaves in de Sitter spacetime and study the implications for causality.  We discuss how causality conditions of AdS and flat spacetime can be applied in de Sitter spacetime, using spatial shifts measured on the boundary to define ``fastest null geodesics" and taking into account the ``stretching" of the de Sitter Penrose diagram.  We consider the propagation of a massless spin-1 field with an $RFF$ coupling in a de Sitter shockwave background as an illustrative example.  We also briefly discuss connections to the ANEC.

\end{abstract}

\newpage

\thispagestyle{empty}
\tableofcontents
\newpage
\setcounter{page}{1}
\setcounter{footnote}{0}

\section{Introduction and Summary}
Causality conditions provide powerful constraints on low energy theories. For example, in \cite{Camanho:2014apa} it was shown how the positivity of the phase shift in eikonal scattering can be used to place constraints on the coefficients of cubic terms in a low energy theory.  This phase shift is known to be related to the asymptotic time delay or advance that a particle experiences when traversing a shockwave geometry or, equivalently, the Shapiro time delay experienced by a particle traversing a black hole \cite{tHooft:1987vrq,Dray:1984ha,Kabat:1992tb}.  A time advance for a single scattering event that is measurable within the regime of validity of the EFT would indicate asymptotic superluminality in the theory.   Thus interactions that allow for such advances can be ruled out or, alternatively, new physics must appear at a scale indicated by the presence of the time advance.

S-matrix observables and asymptotic observables like the eikonal phase shift are particularly useful in identifying superluminality, as the comparison of the lightcone structure in the bulk of two different spacetimes (e.g., one spacetime with a black hole and one without) can be subtle.  For example, while it was argued in \cite{Visser:1998ua} that, in flat spacetime, the ``tipping" of light cones in the presence of matter is always associated with violations of the null energy condition, it was observed in \cite{Gao:2000ga} that such a statement is coordinate-dependent and that there is no unique way to compare light cones in the bulk of two different asymptotically flat spacetimes.  Instead, in \cite{Gao:2000ga} it was proved that, in a spacetime with a timelike conformal boundary satisfying strong causality, the ``fastest null geodesic" connecting two points on the boundary must run entirely along the boundary and not through the bulk.  In other words, the causal structure on the boundary sets the maximum speed for propagation in the bulk.

The application of the theorem of \cite{Gao:2000ga} to asymptotically anti de Sitter spacetimes is obvious.  In flat spacetime, although the theorem doesn't hold, superluminal propagation through the bulk as measured by a boundary observer is nevertheless taken to be a reasonable indication of a ``sick" theory, as it has been argued that it will generically lead to closed time-like curves (see, e.g., \cite{Camanho:2014apa}).  The superluminality that is used as a diagnostic tool in eikonal scattering can be thought of as superluminality defined relative to a lightcone on the boundary. In this paper we consider the extent to which this notion of superluminality can be applied to asymptotically de Sitter spacetimes.  In particular, we wish to understand whether it is possible to diagnose potential pathologies in de Sitter using observables that can be defined unambigously on the future boundary of de Sitter.

Generalizing to de Sitter spacetime is non-trivial for several reasons. There is no de Sitter S-matrix so the positivity of the phase shift in eikonal scattering can't be used.  However, one can still perform the equivalent calculation of the shift experienced by a null ray traversing a black hole or shockwave in curved spacetime.  Curiously, the calculation of the Shapiro time delay in de Sitter gives a negative result at large impact parameter, naively indicating a time advance and thus acausality, even in a free theory in a very simple and obviously physical background.  To actually define the presence of superluminality unambiguously, one should restrict consideration to quantities defined on the boundary, where it is valid to compare observables between two different asymptotically de Sitter spacetimes.  However, the boundary of de Sitter spacetime is spacelike, with no time direction on which to specify a time delay or advance. Furthermore, perturbations in the bulk of de Sitter can distort the Penrose diagram, stretching it out in the time direction \cite{Gao:2000ga}, complicating the interpretation of a time advance. 

In this work we consider the classical calculations of Shapiro time delay due to black holes and, equivalently, shockwaves in de Sitter spacetime.  We argue that fastest null geodesics can be defined unambiguously by considering spatial shifts on the boundary and taking into account the stretching of the de Sitter Penrose diagram.  In particular, while there is of course no fastest null geodesic that lies entirely on the boundary, a straightforward generalization of the flat spacetime and anti de Sitter causality conditions is to require that the fastest null geodesic, measured by the greatest positive spatial shift on the boundary, is the geodesic at largest impact parameter from the matter perturbation.  Stated equivalently, the widest light cone should be that of the null ray at largest impact parameter.  By this criteria, the naive time advance seen in Schwarzschild de Sitter spacetime or in the shockwave background is a measure of the stretching of the overall Penrose diagram rather than an indication of superluminal propagation.   We emphasize that we are not proving a causality condition from more fundamental principles but instead are assuming that, in a healthy theory, light rays shouldn't ``speed up" when traversing a gravitational potential, and we are showing that it is possible to enforce this criteria in de Sitter spacetime in an unambiguous way.  

After considering null geodesics in black hole and shockwave backgrounds, we then consider the propagation of a massless spin-1 particle in the shockwave background.  We compare the case of a minimally coupled photon to a non-minimally coupled photon with an $RFF$ coupling.  While the minimally coupled photon follows the null geodesic as expected, the non-minimally coupled photon at small impact parameter travels on a lightcone that is wider than the lightcone of the photon at largest impact parameter, violating our causality condition.  Furthermore,  while the presence of the black hole or shockwave can make more of the future boundary of de Sitter reachable by a single minimally-coupled observer due to the stretching of the Penrose diagram, this increase is finite, with $\Delta \chi \sim 2GM \times H$ where $\chi$ is the spatial conformal coordinate on the boundary and $H$ is the de Sitter Hubble constant. We find that with the $RFF$ coupling, however, an observer can potentially reach the entire future boundary of de Sitter by passing close enough to the shockwave.  This further implies a pathology of this coupling.

Connections between causality and energy conditions have a long history (see, e.g., \cite{Morris:1988tu,Visser:1998ua,Gao:2000ga,Dubovsky:2005xd,Kelly:2014mra,Engelhardt:2016aoo,Hartman:2016lgu} and references therein).  We finish by briefly commenting on the relation of our considerations to the average null energy condition. In particular, while positivity of the shift experienced by a null ray traversing a shockwave in flat spacetime is directly related to the positivity of the averaged null energy, we see how negative shifts in de Sitter spacetime are consistent with positive average null energy.

\bigskip

\section{Null Geodesics in de Sitter Spacetime}

\begin{figure}[ht]
\begin{subfigure}{.5\textwidth}
  \centering
  \epsfig{file=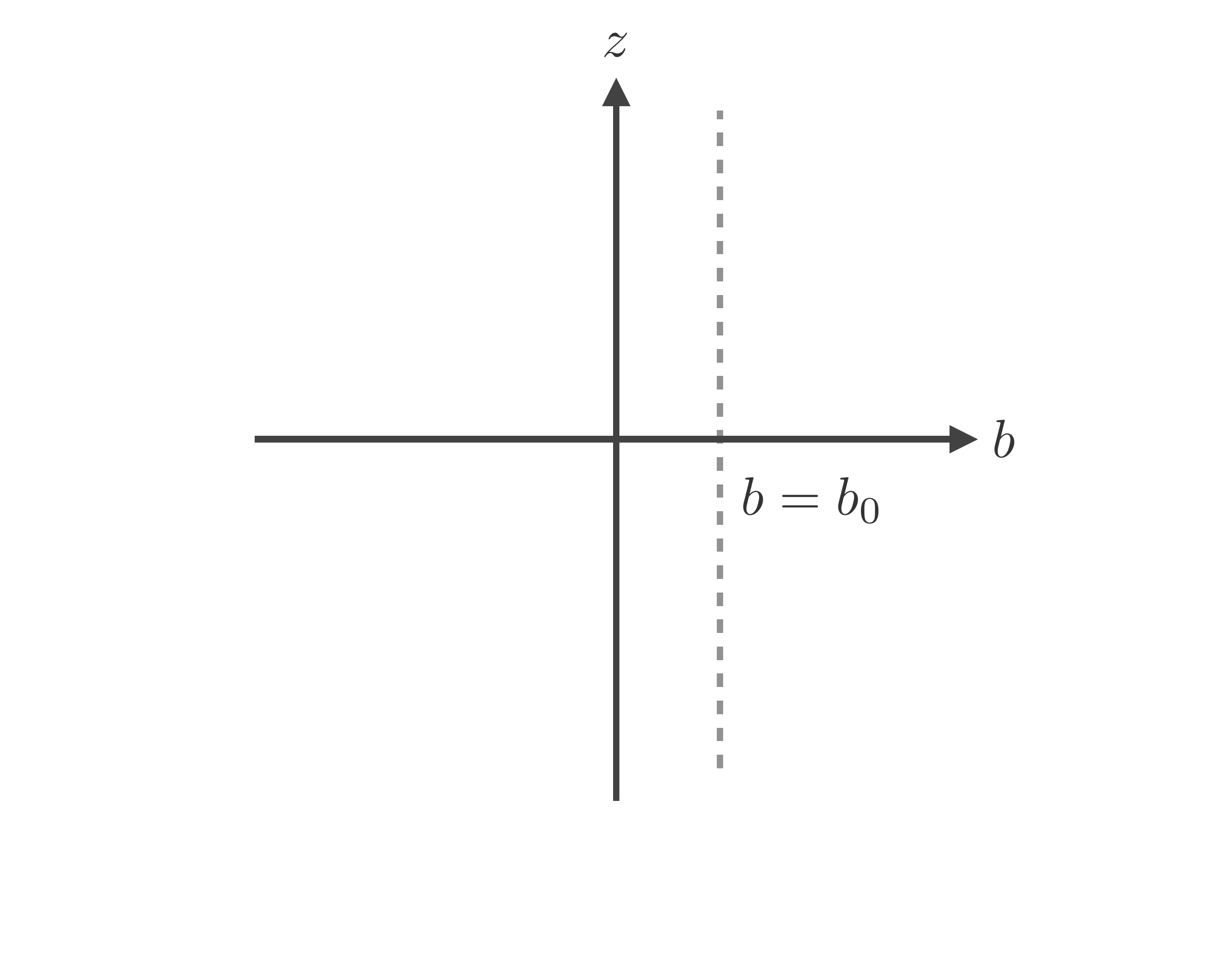,width=1\linewidth}
  \caption{Static Coordinates}
  \label{fig:geostat}
\end{subfigure}
\begin{subfigure}{.5\textwidth}
  \centering
\epsfig{file=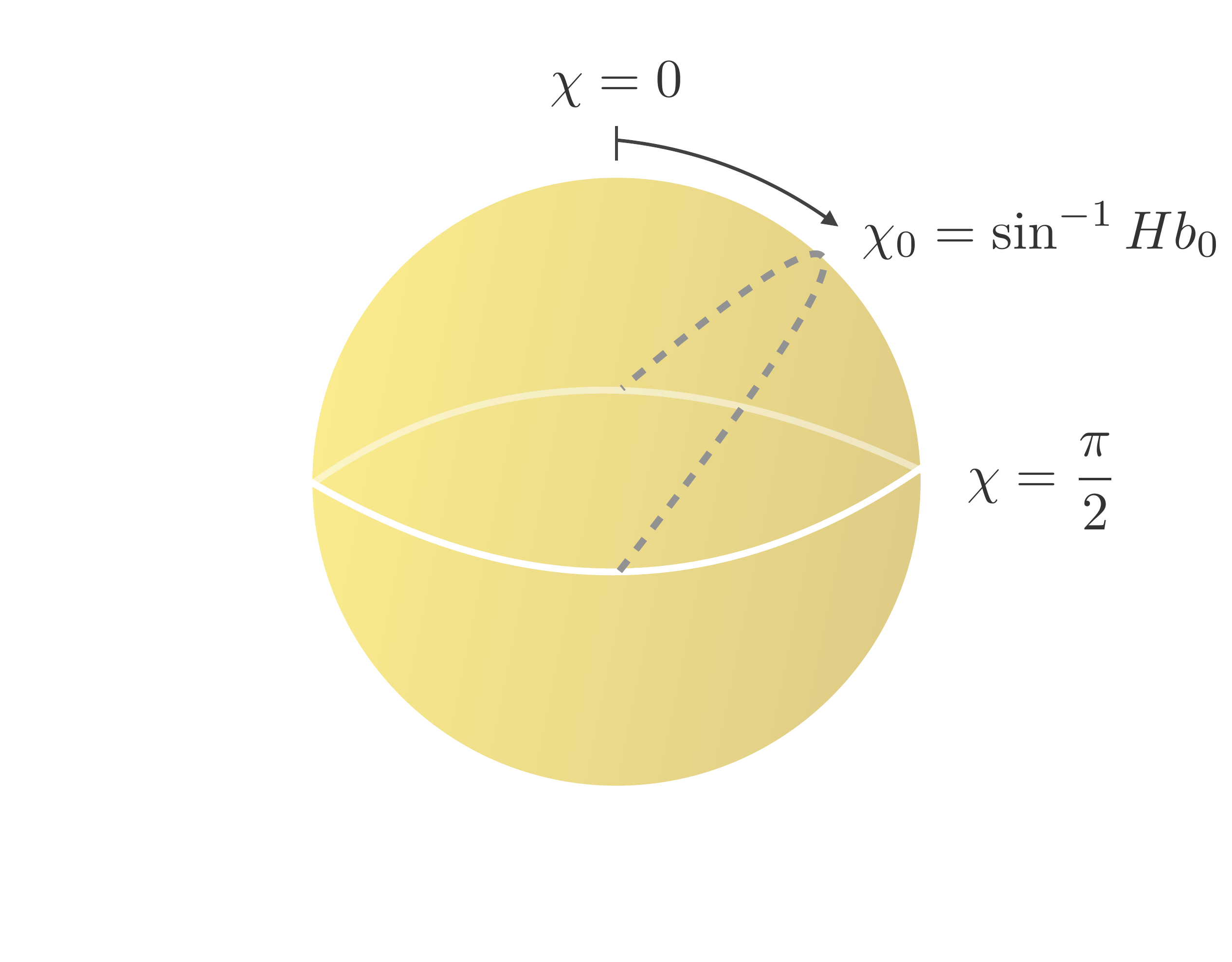,width=1\linewidth} 
  \caption{Conformal Coordinates}
  \label{fig:geoconf}
\end{subfigure}
\caption{In both figures the dashed line represents the spatial trajectory of the same null geodesic at impact parameter $b=b_0$.  This is depicted in static coordinates in (a) and conformal coordinates in (b).}
\label{fig:geodesics}
\end{figure}

In this work we will consider the path of null geodesics through the bulk of asymptotically de Sitter spacetimes, both in empty de Sitter and in the presence of matter.  We will use two coordinate systems, depending on convenience: static coordinates
\be
\label{static}
ds^2 = -\left(1-H^2\,r^2\right)dt^2 +\frac{dr^2}{1-H^2\,r^2} +r^2 d\Omega^2\, ,
\ee
 and conformal coordinates
\be
\label{conf}
ds^2 =\frac{1}{\cos^2HT}  \left(-dT^2 +\frac{1}{H^2}d\chi^2+\frac{1}{H^2}\sin^2\chi \, d\Omega^2\right) \, ,
\ee
where
\be
-\frac{\pi}{2} \leq HT \leq \frac{\pi}{2} \, , ~~~~ {\rm and} ~~~~ 0 \leq \chi \leq \pi \, .
\ee
The conversion between the two is given by the coordinate transformation
\bea
\tanh Ht =\left\{
                \begin{array}{ll}
                  \frac{\sin HT}{\cos \chi} ~~\mbox{for} ~~ Hr < 1  \\
                  \\
                  \frac{\cos \chi}{\sin HT} ~~\mbox{for} ~~ Hr > 1
                \end{array}
              \right. 
~~~~{\rm and}~~~~              
Hr = \frac{\sin \chi}{\cos HT} \, .
 \eea
In this way, we will use the static coordinates outside of the static patch (i.e., for $H r>1$), with $t$ becoming a spatial coordinate and $r$ becoming a time coordinate.

In static coordinates we introduce cylindrical coordinates so that $r^2 = b^2 +z^2$.  We will consider null geodesics traveling in the $z$-direction at constant impact parameter $b=b_0$.  In empty de Sitter, their trajectory in the $z$-direction is given by
\bea
\tanh Ht =\left\{
                \begin{array}{ll}
                 \frac{H z}{\sqrt{1-H^2 b_0^2}} ~~\mbox{for} ~~ Hr < 1  \\
                  \\
                \frac{\sqrt{1-H^2 b_0^2}}{H z}~~\mbox{for} ~~ Hr > 1
                \end{array}
              \right. \, .
\eea
As $r$ goes from $b_0$ to $\frac{1}{H}$, $t$ goes from $0$ to $+ \infty$ and as $r$ goes from $\frac{1}{H}$ to $\infty$, $t$ goes back from $+ \infty$ to $0$.

\begin{figure}[t!]
\begin{center}
\epsfig{file=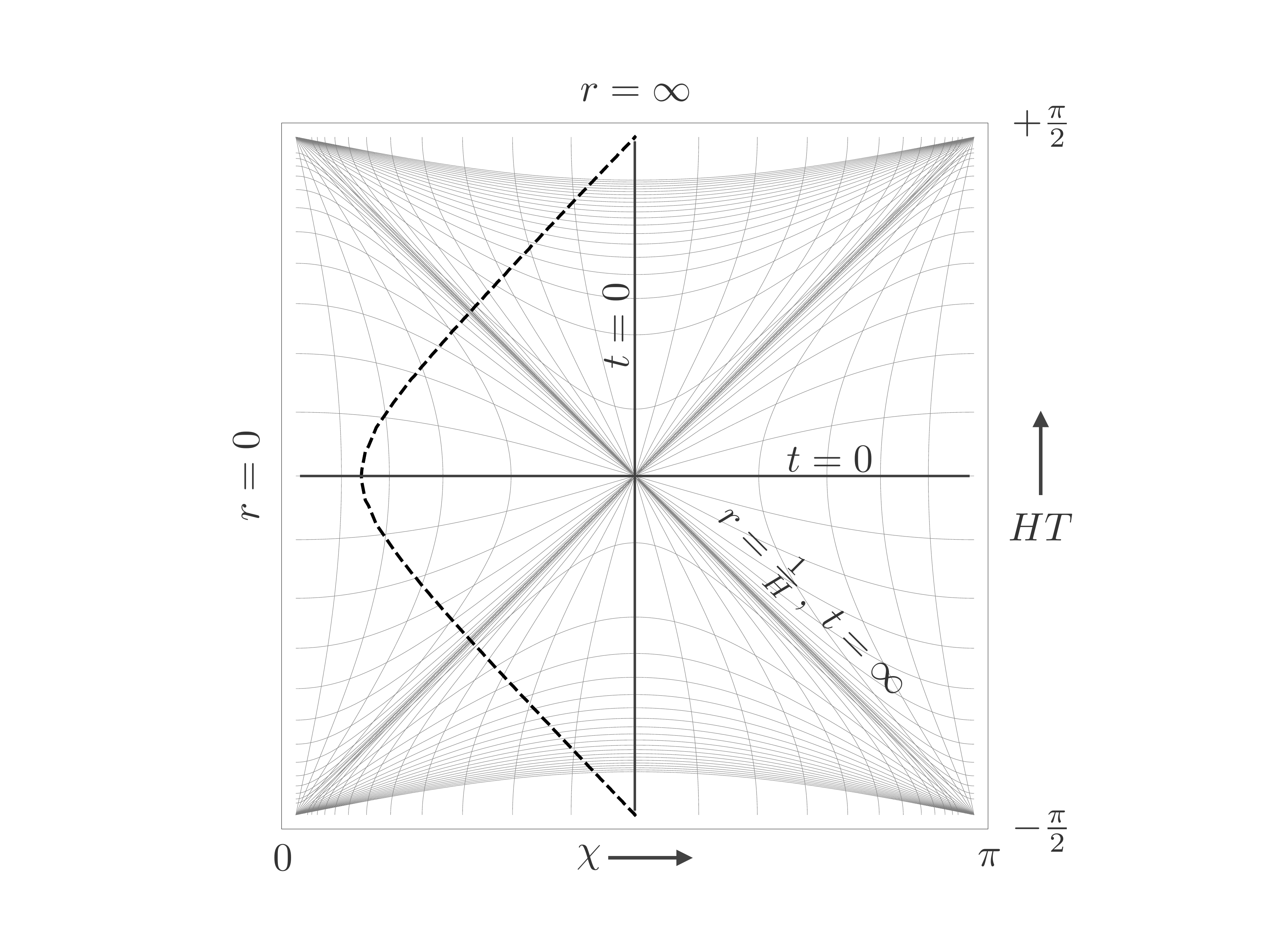,width=6.0in}
\caption{\small The dashed line represents a null geodesic at finite impact parameter $b=b_0$ on the de Sitter Penrose diagram.}
\label{fig:penrose}
\end{center}
\end{figure}

In conformal coordinates the same null geodesics are given by
\be
\frac{\cos \chi}{\cos HT}=\sqrt{1-H^2b_0^2} = \cos \chi_0 \, ,
\ee
where 
\be
\sin \chi_0 = H b_0 \, .
\ee
These are simply great circles on the de Sitter sphere.  As conformal time $T$ ranges from $-\frac{\pi}{2}$ to $\frac{\pi}{2}$, $\chi$ goes from $\frac{\pi}{2}$ to $\chi_0$ and back to $\frac{\pi}{2}$.  In Figure \ref{fig:geodesics} we plot the spatial trajectory of the same null geodesic in static and conformal coordinates.  In Figure \ref{fig:penrose} we show the trajectory of the null geodesic at finite impact parameter in the Penrose diagram of empty de Sitter spacetime.

In what follows we will consider how the presence of matter in the bulk of de Sitter spacetime effects these null geodesics.  In particular, we will consider the case of a Schwarzschild de Sitter black hole and, equivalently, a shockwave.  Our interest will be in how the presence of matter in the bulk changes where the null geodesic hits the future boundary of de Sitter spacetime, i.e., ${\cal I}^+$.  We will use these shifts to define a notion of the ``fastest null geodesic."

\bigskip

\section{Shapiro Time Delay}
\subsection{Coordinate Shift}
We first consider the effect of a black hole located at $r=0$ on the trajectory of a null geodesic at impact parameter $b=b_0$.  We are interested in the resulting spatial shift $\Delta \chi$ on the future boundary of de Sitter spacetime.  We start in static coordinates so that the Schwarzschild de Sitter metric is of the usual form
\be
\label{SdS}
ds^2 = -(1-\frac{2 G M}{r} -H^2r^2)dt^2+\frac{dr^2}{1-\frac{2 G M}{r} -H^2r^2}+r^2 d\Omega^2 \, .
\ee
Ignoring any deflection in the transverse direction due to the black hole, we consider a null ray traveling at impact parameter $b_0$ so that $r^2 = b_0^2+z^2$.\footnote{One can imagine a second black hole placed symmetrically at $b=2 b_0$ to achieve this.}  We work perturbatively in $2GM$ and we assume $b_0 \gg 2 G M$ to avoid the black hole horizon.  Setting $ds^2 =0$ we can solve for the shift in the $t$-coordinate as a function of $z$, to leading order in $2 GM$:\footnote{See, e.g., \cite{Camanho:2014apa} for the analogous calculation in flat spacetime.}
\be
\label{dtdS}
\Delta t = \frac{2GM}{\sqrt{1-H^2b_0^2}} \int dz \, \frac{2z^2+b_0^2\left(1-H^2(b_0^2+z^2)\right)}{2(b_0^2+z^2)^{\frac{3}{2}}  \left( 1-H^2(b_0^2+z^2)\right)^2} \, .
\ee
We note that the ``perturbation" blows up near $r=\tfrac{1}{H}$ and thus, naively, our expansion breaks down here.  However, all terms in the $2 G M$ expansion come with the same power of $1/(1-H^2 r^2)$. An expansion in $2GM$ is therefore valid: one can simply pull out a factor of $1/(1-H^2 r^2)^2$ in front of the series.  Thus $\Delta t$ can be used reliably to calculate the shift as $r \rightarrow \infty$.  We find:
\be
\Delta t_{\rm in} =
2\, GM \left(\tanh^{-1}\left[\frac{z}{\sqrt{(1-H^2 b_0^2)(b_0^2+z^2)}} \right] 
-\frac{z(1-2H^2(b_0^2+z^2))}{2\sqrt{(1-H^2b_0^2)(b_0^2+z^2)}(1-H^2(b_0^2+z^2))}
\right) \, ,
\ee
for $Hr<1$ and
\be
\Delta t_{\rm out} =
2\, GM \left(\tanh^{-1}\left[\frac{\sqrt{(1-H^2 b_0^2)(b_0^2+z^2)}}{z} \right] 
-\frac{z(1-2H^2(b_0^2+z^2))}{2\sqrt{(1-H^2b_0^2)(b_0^2+z^2)}(1-H^2(b_0^2+z^2))}
\right)\, ,
\ee
for $Hr >1$.  Here we've fixed the trajectory so that it always passes closest to the black hole at conformal time $T=0$.  This way the path of the light ray is symmetric about the $T=0$ axis of the Penrose diagram.

We emphasize that the behavior of $\Delta t$ in the bulk on its own is not particularly meaningful as it is highly coordinate dependent, including the apparent divergence at $Hr=1$. However, we can use this expression to determine the shift of the light ray on the future boundary of de Sitter, given in these coordinates by $r \rightarrow \infty$.  In the limit that $r \rightarrow \infty$, the shift in the $t$-coordinate becomes
\be
\label{tdS}
\Delta t_{dS} =  
2\,GM \left(\log\left[\frac{1}{H b_0}+\sqrt{\frac{1}{H^2b_0^2}-1} \right] -\frac{1}{\sqrt{1-H^2b_0^2}} \right) \,.
\ee
In the next section we will use this expression to determine the spatial shift on the future boundary of de Sitter and the implications for the de Sitter Penrose diagram.

\subsection{Boundary Shift and Penrose Diagrams}
\label{rot}
The maximum possible impact parameter is $H b_0=1$: beyond this, the geodesic enters the southern hemisphere of the de Sitter sphere and starts getting closer to the black hole located at $\chi=\pi$.  Indeed, one of the features of de Sitter spacetime that makes it challenging to define asymptotic superluminalty is that asymptotic distances are temporal rather than spatial.

In the setup being considered above, $\Delta t$ naively blows up as $H b_0$ approaches $1$.  However, this is simply an artifact of our choice of coordinates.  Let us perform a spatial rotation (i.e., a de Sitter isometry) so that the black hole is located at $b=-b_0$ and the null ray passes through $b=0$ instead.  It is easiest to do so in embedding space coordinates $(X^0,X^1,X^2,X^3,X^4)$:
\begin{align}
    & X^0 = \tfrac{1}{H}\sqrt{1-H^2 r^2}\, \sinh Ht ~{\rm for}~ H r<1 \, , ~{\rm and}~=\tfrac{1}{H}\sqrt{1-H^2 r^2}\,  \cosh Ht ~{\rm for}~ H r>1 \, , \nonumber \\
    & X^1 = r\, \cos \theta\, , \nonumber\\
    & X^2 = r\, \sin\theta \cos \phi \, , \\
    & X^3 = r\, \sin\theta \sin \phi \, ,\nonumber \\
    & X^4 = \tfrac{1}{H}\sqrt{1-H^2 r^2}\,  \cosh Ht ~{\rm for}~ H r<1 \, ,~
    {\rm and}~=\tfrac{1}{H}\sqrt{1-H^2 r^2}\,  \sinh Ht ~{\rm for}~ H r>1\, . \nonumber
\end{align}
Expressing the Schwarzschild de Sitter metric \eqref{SdS} in these coordinates,  a rotation through the angle $\chi_0 = \sin^{-1} Hb_0$ can be achieved via the transformation 
\be
X^2 \rightarrow \sqrt{1-H^2b_0^2} \,X^2 + H b_0\, X^4 \, , ~~~~
X^4 \rightarrow  H b_0\,X^2 - \sqrt{1-H^2b_0^2}\, X^4 \, .
\ee

\begin{figure}[t]
\begin{subfigure}{.5\textwidth}
  \centering
  \epsfig{file=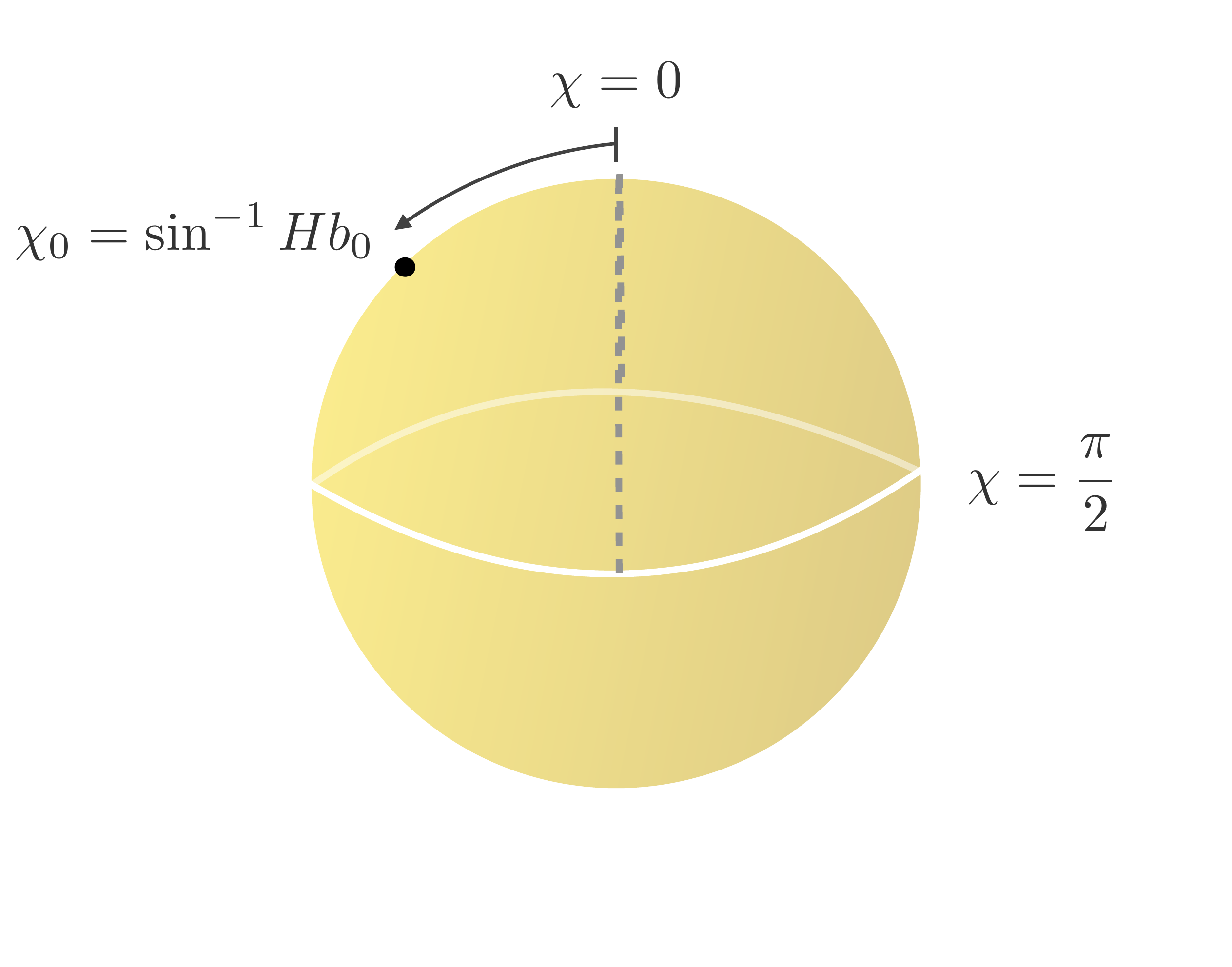,width=1\linewidth}
  \caption{Rotated Black Hole and Null Geodesic}
  \label{fig:rot}
\end{subfigure}
\begin{subfigure}{.5\textwidth}
  \centering
\epsfig{file=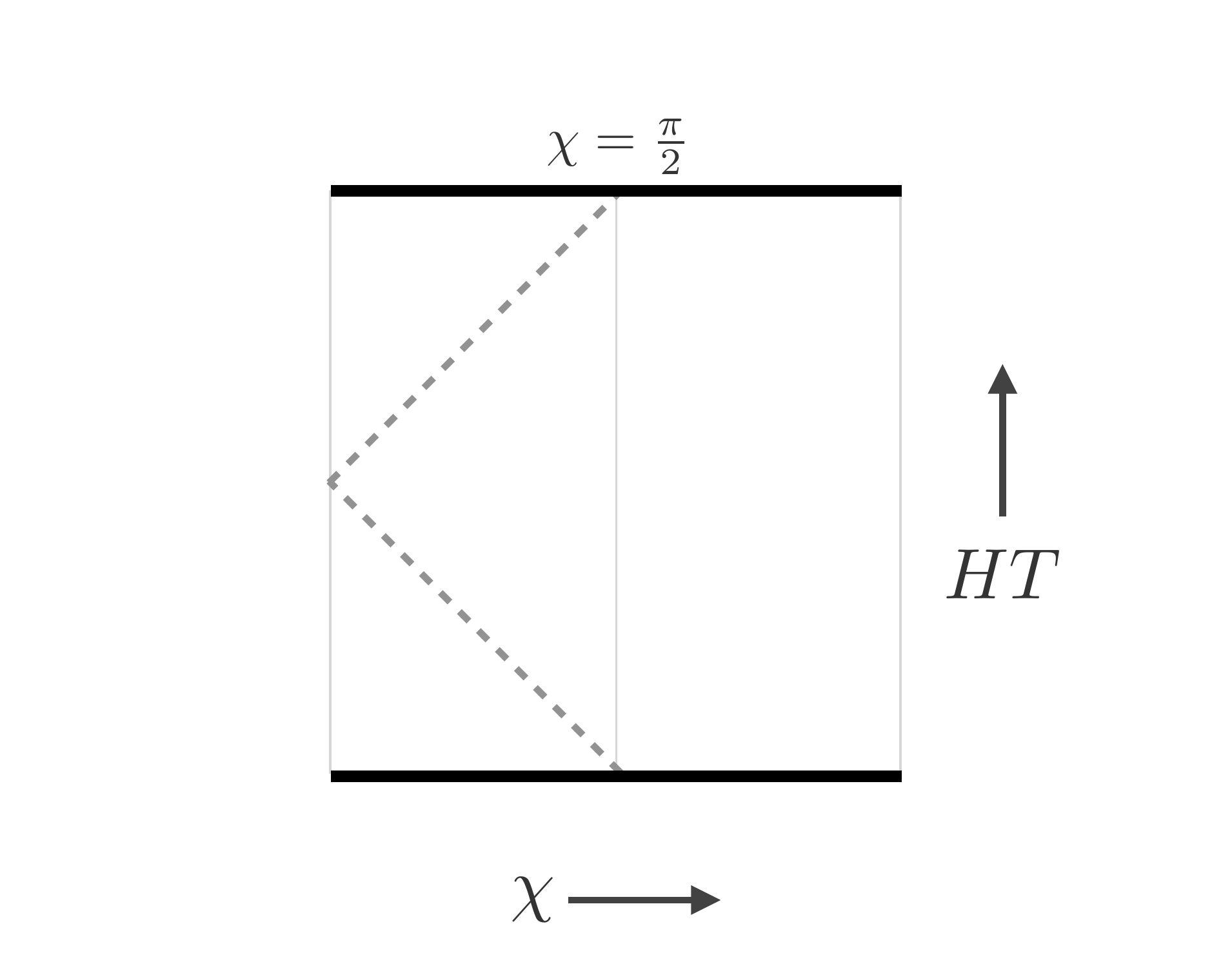,width=1\linewidth} 
  \caption{Penrose Diagram}
  \label{fig:Prot}
\end{subfigure}
\caption{The trajectory of the relevant null geodesic in de Sitter spacetime after rotating. We show the trajectory of this rotated geodesic on the Penrose diagram for empty de Sitter spacetime in (b) where it now travels at a 45 degree angle in the $(T,\chi)$ plane.}
\label{fig:rotgeo}
\end{figure}

We depict the rotated trajectory of the null geodesic passing by the black hole in Figure \ref{fig:rot}.  Importantly, after performing this rotation, light rays now propagate at 45 degree angles in the $(T,\chi)$ Penrose diagram, depicted in Figure \ref{fig:Prot}.  Furthermore, repeating the calculation of the previous section for the rotated metric, the shift in $\Delta t$ acquires an overall factor of $\sqrt{1-H^2b_0^2}$:
\be
\label{rotated}
\Delta t_{dS} \rightarrow  
2\,GM  \left(\sqrt{1-H^2b_0^2}\,\log\left[\frac{1}{H b_0}+\sqrt{\frac{1}{H^2b_0^2}-1} \right] -1 \right) \,.
\ee
We can compare this shift to the result in flat spacetime
\be
\label{flat}
\Delta t_{\rm flat} = 2\, G M \left(\log \left[ \frac{z_0}{b_0}+\sqrt{\frac{z_0^2}{b_0^2}+1}\, \right] -\frac{1}{2\sqrt{1+\frac{b_0^2}{z_0^2}}}\right) \,,
\ee
where $z_0$ is an IR cutoff, $z_0 \gg b_0$.  In flat spacetime, this shift is always positive so that the shift of the time coordinate at the boundary is always a time delay rather than a time advance.  The implication is that this is an indication of a causal theory: light through the bulk propagates more slowly than light on the boundary.  The causal structure of the spacetime is set by the boundary theory.

\begin{figure}[t]
\begin{subfigure}{.5\textwidth}
  \centering
  \epsfig{file=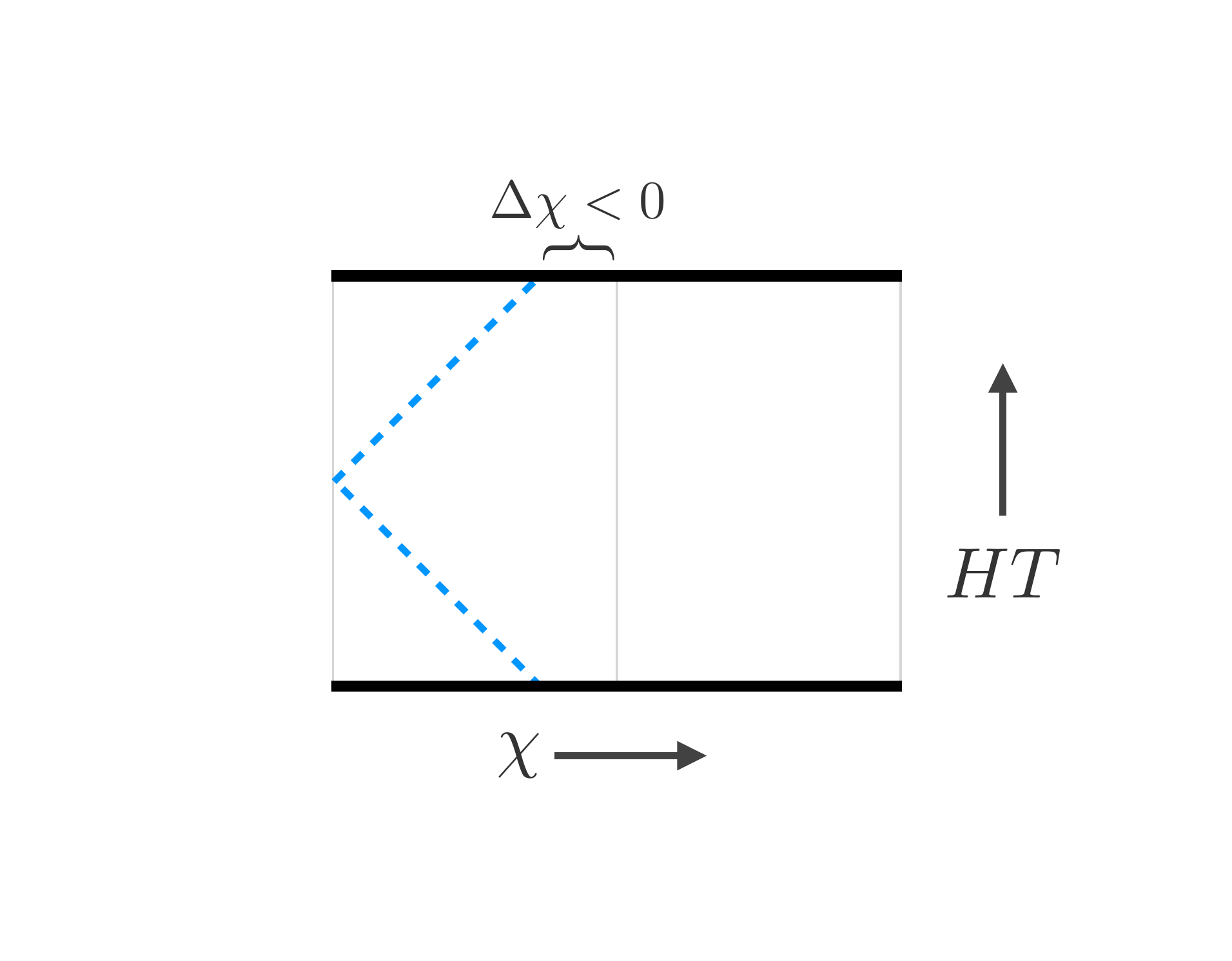,width=1\linewidth}
  \caption{Penrose diagram for $H b_0 < 0.552$}
  \label{fig:shrink}
\end{subfigure}
\begin{subfigure}{.5\textwidth}
  \centering
\epsfig{file=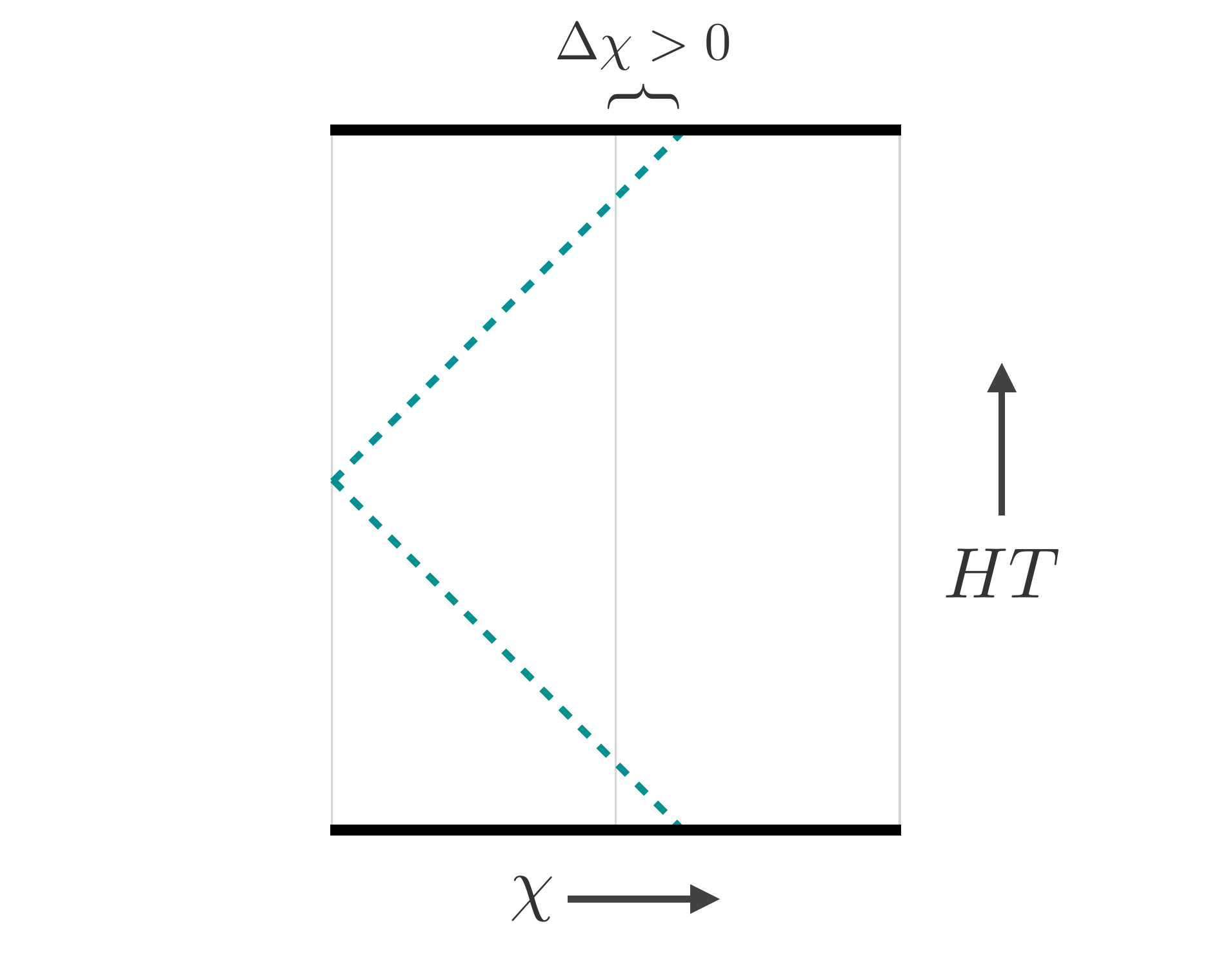,width=1\linewidth} 
  \caption{Penrose diagram for $H b_0 > 0.552$}
  \label{fig:stretch}
\end{subfigure}
\caption{The distortion of the Penrose diagram as seem by a light ray passing by a black hole at different impact parameters.  The larger the impact parameter, the greater the ``stretching" of the diagram.}
\label{fig:shrinkstretch}
\end{figure}

In contrast, in de Sitter spacetime we see that the shift in the $t$-coordinate can be negative for large enough impact parameter $b_0$.  In particular, solving for the root of the above expression \eqref{rotated}, the shift changes from positive to negative when $H b_0 = 0.552434\ldots$ and then becomes increasingly negative as $b_0$ increases.  However, unlike in flat spacetime, this shift doesn't have the interpretation of a time delay or advance on the boundary.  It is a spatial shift instead.  The change in the conformal coordinate $\chi$ at the future boundary is given by
\be
\cos (\chi+\Delta \chi )= \tanh [H (t+\Delta t)] \, ,
\ee
or
\be
\Delta \chi \simeq - H\Delta t \, ,
\ee
for small $\Delta t$.  In other words, a negative value for $\Delta t$ indicates a greater distance traveled, i.e., positive shift at the boundary for $\Delta \chi$.  In particular, if we consider the Penrose diagram, the shift tells us that a ``right"-moving light ray arrives at the future boundary shifted to the left for $H b_0 < 0.552434\ldots$ and is shifted to the right for $Hb_0 > 0.552434\ldots$.  In addition, there is a maximum finite shift at $Hb_0 =1$ given by
\be
\label{finite}
\Delta \chi_{\rm max} = H \times 2 \, GM \, .
\ee
These left/right shifts on the boundary can be depicted as distortions of the Penrose diagram.  Fixing light rays to travel 45 degree angles, a negative shift in $\Delta \chi$ indicates a shrinking of the Penrose diagram while a positive $\Delta \chi$ indicates a stretching instead.  We depict this in Figure \ref{fig:shrinkstretch}.

\begin{figure}[t]
\begin{center}
\epsfig{file=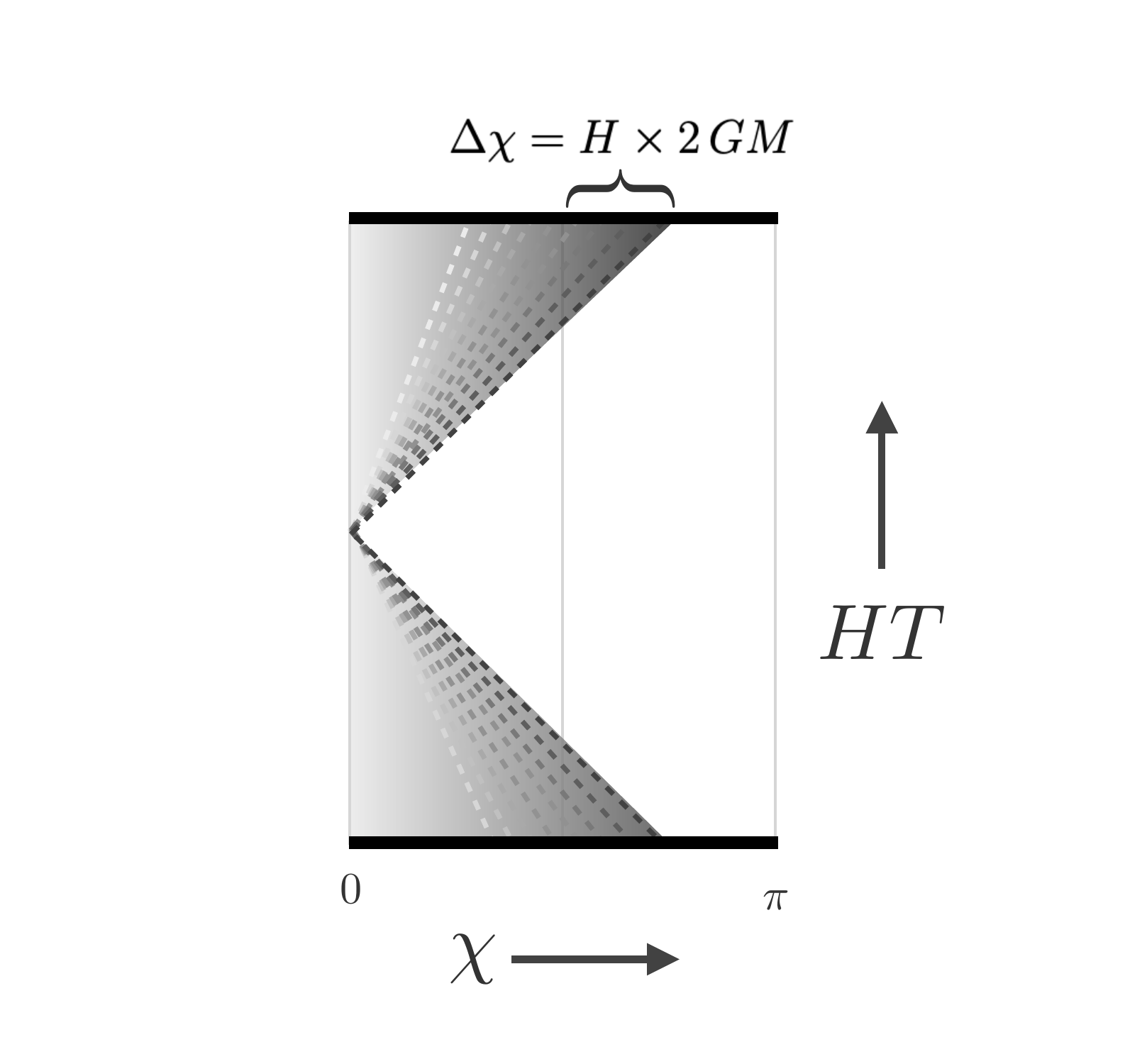,width=4.0in}
\caption{\small Using the null geodesic at largest impact parameter to set the causal structure.  All other null geodesics at smaller impact parameter lie within this light cone.}
\label{fig:maxstretch}
\end{center}
\end{figure}

In de Sitter spacetime, there is no fiducial null geodesic living on the boundary that we can compare with the null geodesics traveling through the bulk.  However, we can instead consider the light ray at maximum impact parameter from the black hole $H b_0 =1$ as our fiducial null geodesic, as it is the farthest from the source and can be defined in a coordinate independent way.  If we use this geodesic to set the causal structure of the spacetime in the presence of the black hole, we see that all other light rays passing closer to the black hole travel within the light cone of this geodesic.  I.e., the null geodesic at largest impact parameter has traveled the farthest in the same amount of conformal time $T$ as the other light rays traversing the space.  We would thus conclude that the ``fastest" null geodesics are, unsurprisingly, those at largest impact parameter from the black hole and that we can't ``speed up" a light ray by sending it closer to a black hole.  We depict this in Figure \ref{fig:maxstretch}.

We note that in de Sitter we can think of there being two competing effects: the slowing of light rays by a source like a black hole and the stretching of the Penrose diagram which persists even at the maximum impact parameter.  For the light ray at impact parameter $H b_0 = 0.552434$, these two effects precisely cancel and give zero shift.  By using the light ray at maximum impact parameter to set the causal structure, we can disentangle the sign of the shift from causality considerations.

Finally, we note that while the presence of the black hole has made more of the future boundary of de Sitter accessible to an observer who passes through $\chi = 0$ at $T=0$, it is only a finite amount, given by equation \eqref{finite}.  This is in contrast to what we will find for the non-minimally coupled photon in section \ref{RFF}, where the entire future boundary will become accessible at sufficiently small impact parameter.

\bigskip

\section{Shockwaves}
The total shift experienced by a particle traversing a black hole is equal to that experienced by a particle traversing a shockwave as the two are related by an ultrarelativistic boost \cite{Aichelburg:1970dh}. In this section we briefly verify this explicitly.  This calculation will also be useful in the following section where we consider the propagation of interacting massless particles in the shockwave background.  Shockwave solutions in de Sitter spacetime were first found in \cite{Hotta:1992qy} and our results are consistent with those solutions.

To study the shockwave, we introduce lightcone coordinates $u,v$ in the $t-z$ plane of the static coordinates:
\be
\label{ttou}
t= \frac{1}{2 H} \log \left[ \frac{(1+Hu )(1+Hv)}{(1-Hu)(1-Hv)}   \right] \, , ~~~~ z = \sqrt{1-H^2b^2}\,\frac{v-u}{1-H^2u v} \, .
\ee
The de Sitter metric becomes
\be
\label{pen}
ds_{\rm dS}^2 =-4\left(1-H^2 b^2 \right) \frac{du\,dv}{\left(1-H^2 u v \right)^2}+\frac{db^2}{1-H^2 b^2}+b^2 d\phi^2 \, .
\ee
We now adopt a Kerr-Schild ansatz for the metric in the presence of a shockwave so that metric becomes
\be
\label{dS}
ds^2 = ds_{\rm dS}^2 + \Delta ds^2 \, ,
\ee
with the metric perturbation given by
\be
\label{lc}
\Delta ds^2 =  F[x] (k_\mu dx^\mu)^2\, ,
\ee
where $k_\mu= (0,-1,\vec{0})$.   The general equations of motion can be written as
\be
\Box h^\mu_{~\nu}-2 H^2 h^\mu_{~\nu} = -16 \pi G\, T^\mu_{~\nu} \, ,
\ee
where $h^\mu_{~\nu}\equiv F[v,\vec{x}]k^\mu k_\nu$.  We take the source to be that of a point particle moving at the speed of light with momentum $p_u$:
\be
T_{\mu\nu}=p_u\,\delta(v)\delta(b) k_\mu k_\nu \, .
\ee
Taking $F[v,\vec{x}] \rightarrow 2Gp_u\,f(b)\,\delta(v) $, the equations of motion become
\be
\left(1-H^2b^2\right)f''(b)+\frac{1}{b}f'(b)+2H^2\left(1+\frac{1}{1-H^2 b^2} \right)f(b)= -8  \pi \,\delta(\vec{x}) \, .
\ee
The solution is given by
\be
\label{Fb}
f(b) =  4\,\sqrt{1-H^2b^2}\,\, \left(\sqrt{1-H^2b^2}\,\log\! \left[\frac{1}{H b}+\sqrt{\frac{1}{H^2b^2}-1}\right] -1 \right)  \, .
\ee
This solution agrees with that obtained in \cite{Hotta:1992qy} with the appropriate pullback from embedding space coordinates, i.e., with the identification $Hb = \sin \theta$.

We now consider the shift in the $u$-coordinate experienced by a null geodesic traveling in the $v$ direction at impact parameter $b_0$ and crossing the shockwave at $v=0$. Setting $ds^2 =0$ for the metric \eqref{dS} and solving for $du$, we see that the shift is given by
\be
\label{delu}
\Delta u =2Gp_u\frac{f(b_0)}{4(1-H^2 b_0^2)}\,  \, .
\ee
We can relate the shift $\Delta u$ to the time shift $\Delta t$ calculated for the Shapiro time delay in the previous section.  Using \eqref{ttou} at $v=0$ and expanding in small $\Delta u$, gives
\be
\Delta t \simeq \Delta u= 2 G p_u \left(\log\left[\frac{1}{H b_0}+\sqrt{\frac{1}{H^2b_0^2}-1} \right] -\frac{1}{\sqrt{1-H^2b_0^2}} \right) \,.
\ee
Identifying $2GM$ = $2Gp_u$, we see agreement with \eqref{tdS}.  Again we note the negative values of $\Delta u$ at large impact parameter.

\begin{figure}[t]
\begin{subfigure}{.5\textwidth}
  \centering
  \epsfig{file=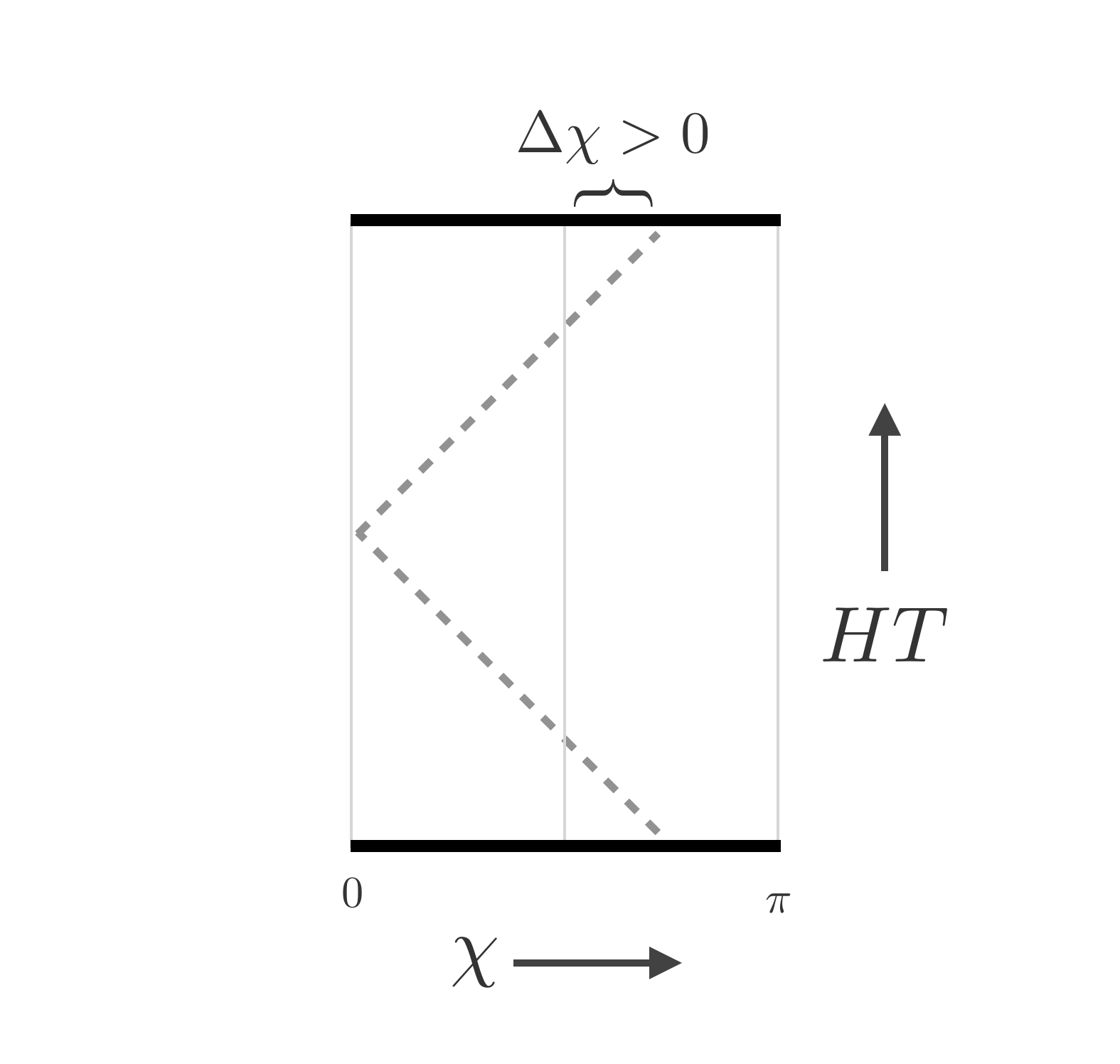,width=1\linewidth}
  \caption{Shockwave Penrose diagram}
  \label{fig:shock1}
\end{subfigure}
\begin{subfigure}{.5\textwidth}
  \centering
\epsfig{file=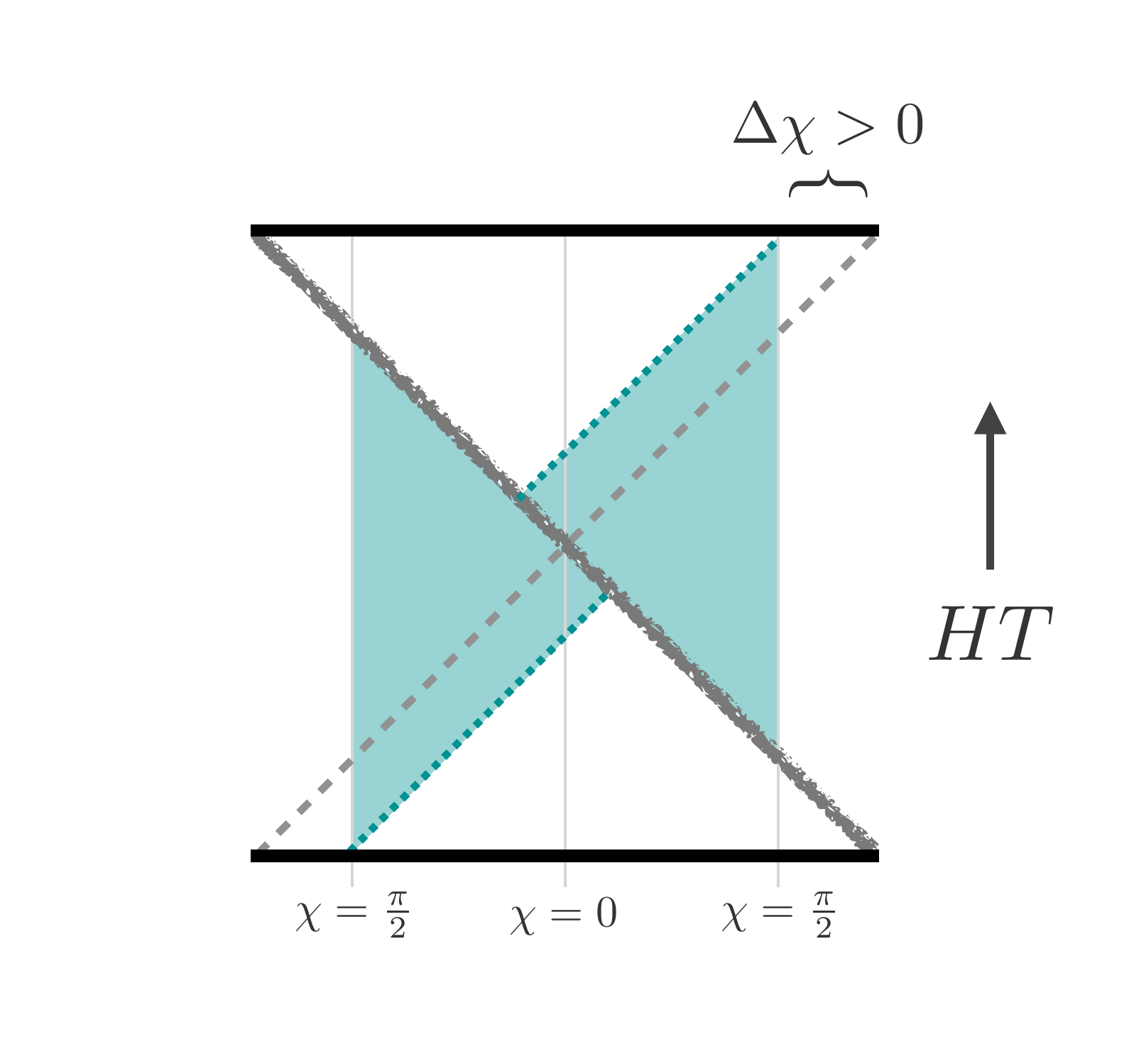,width=1\linewidth} 
  \caption{``Unfolded" shockwave Penrose diagram}
  \label{fig:shock2}
\end{subfigure}
\caption{The distortion of the Penrose diagram as seen by a light ray traversing a shockwave geometry at large impact parameter.  On the right hand side we've ``unfolded" the diagram to better depict the shock (grey solid line), the null geodesic (grey dashed line) and the causal diamonds of empty de Sitter (teal) that have been brought into causal contact in the presence of the shockwave. }
\label{fig:shock}
\end{figure}

Once again we can rotate our system so that the null geodesic passes though $b=0$ and the shockwave travels past at impact parameter $b=b_0$.  As in the previous section, the result is an overall factor of $\sqrt{1-H^2b_0^2}$ multiplying expression \eqref{delu}.  In Figure \ref{fig:shock} we show the Penrose diagram for the de Sitter shockwave spacetime for a light ray at large impact parameter in this setup.  We see that, while the shockwave does connect regions of the Penrose diagram that were previously out of causal contact \cite{Gao:2000ga}, the negative shift $\Delta u$ can't readily be interpreted as a superluminality relative to some fiducial lightcone. Rather, as in the black hole calculation, the negative shift at large impact parameter is indicating the lengthening of the Penrose diagram.  Again, all null geodesics at smaller impact parameter lie inside the lightcone set by the null geodesic at largest impact parameter, indicating no ``speedup" due to propagation near the shock.

\section{Massless Spin-1 Fields}
\label{RFF}
We now go beyond the geodesic calculation and calculate the phase shift for a massless spin-1 field propagating in a de Sitter shockwave background.  We first consider the minimally coupled massless spin-1 and find a shift that agrees with the geodesic calculation, as expected.  We then consider a non-minimally coupled massless spin-1 field and find that, at small impact parameter, the two polarizations have phase shifts of the opposite sign, indicating a widened lightcone for one of them relative to the lightcone at large impact parameter.  As in flat spacetime, at sufficiently small impact parameter, this negative shift can be made arbitrarily large.

\subsection{Minimal Coupling}
We start by considering a massless spin-1 field minimally coupled to gravity:
\begin{equation}
    S = \int d^{4}x\sqrt{-g}\Big(-\frac{1}{4}F^{\mu \nu}F_{\mu \nu}\Big) \, ,
\end{equation}
with equations of motion
\begin{equation}
  \nabla^{\mu}F_{\mu \nu} = 0 \, .
\end{equation}
Using the coordinate system \eqref{pen}, we perform a decomposition of the spin-1 field into harmonics $Y_{km}(b,\phi)$ on the transverse space (see Appendix \ref{SVT}).  We write
\begin{equation}
\label{decomp}
    A_{\mu}dx^{\mu} = A_{u}(u, v)Y(b,\phi)du + A_{v}(u, v)Y(b,\phi)dv + \Big[B(u, v)D_{I}Y(b,\phi) + C(u, v)V_{I}(b,\phi)\Big]dx^{I}, 
\end{equation}
where $I = {b,\phi}$.  In what follows we will consider the $m=0$ harmonic. We note that, for $m = 0$, $B(u,v)$ corresponds to the polarization in the $\vec{b}$ direction, whereas $C(u,v)$ corresponds to the polarization in the azimuthal direction.

After eliminating $A_u$ and $A_v$, the equations of motion for the physical polarizations in the shockwave metric background are given by
\begin{align}
    &\partial_u\partial_vB(u,v)+\frac{H^2(1+k^2)}{(1 - H^2 uv)^2}B(u,v)  + 2Gp_u\,\delta(v)\frac{f(b)}{4(1 - H^2 b^2)}\partial_{u}^2B(u,v) =0 \, ,\nonumber\\
    & \partial_u\partial_vC(u,v)+\frac{H^2(1+k^2)}{(1 - H^2 uv)^2}C(u,v) + 2Gp_u\,\delta(v)\frac{f(b)}{4(1 - H^2 b^2)}\partial_{u}^2C(u,v) =0\, ,
\end{align}
with $f(b)$ given by \eqref{Fb}.  We note that each polarization obeys the same equation.  Assuming that $B$ and $C$ take the form of a plane wave near the shock, e.g., $B(u, v)\sim e^{ip_v u}B(v)$, these equations are straightforward to solve across the shock:
\begin{equation}
    B(u, v) \sim \exp\Big(ip_v u  + 2i G p_up_v \Theta(v)\frac{f(b)}{4(1 - H^2 b^2)} + ip_u v \Big) \, ,
\end{equation}
where we have taken
\be
H^2(1+k^2) = p_u p_v \,.
\ee
As expected, there is a phase shift in the $u$-coordinate which exactly matches the geodesic analysis.  Multiplying through by the rotation factor $\sqrt{1 - H^2 b^2}$ found in the previous sections, we have
\begin{equation}
    \Delta u =2G p_u\, \frac{f(b)}{4\sqrt{1 - H^2 b^2}} \, .
\end{equation}
Again, we note that the shift is the same for the two polarizations.

\begin{figure}[t]
\begin{subfigure}{.5\textwidth}
  \centering
  \epsfig{file=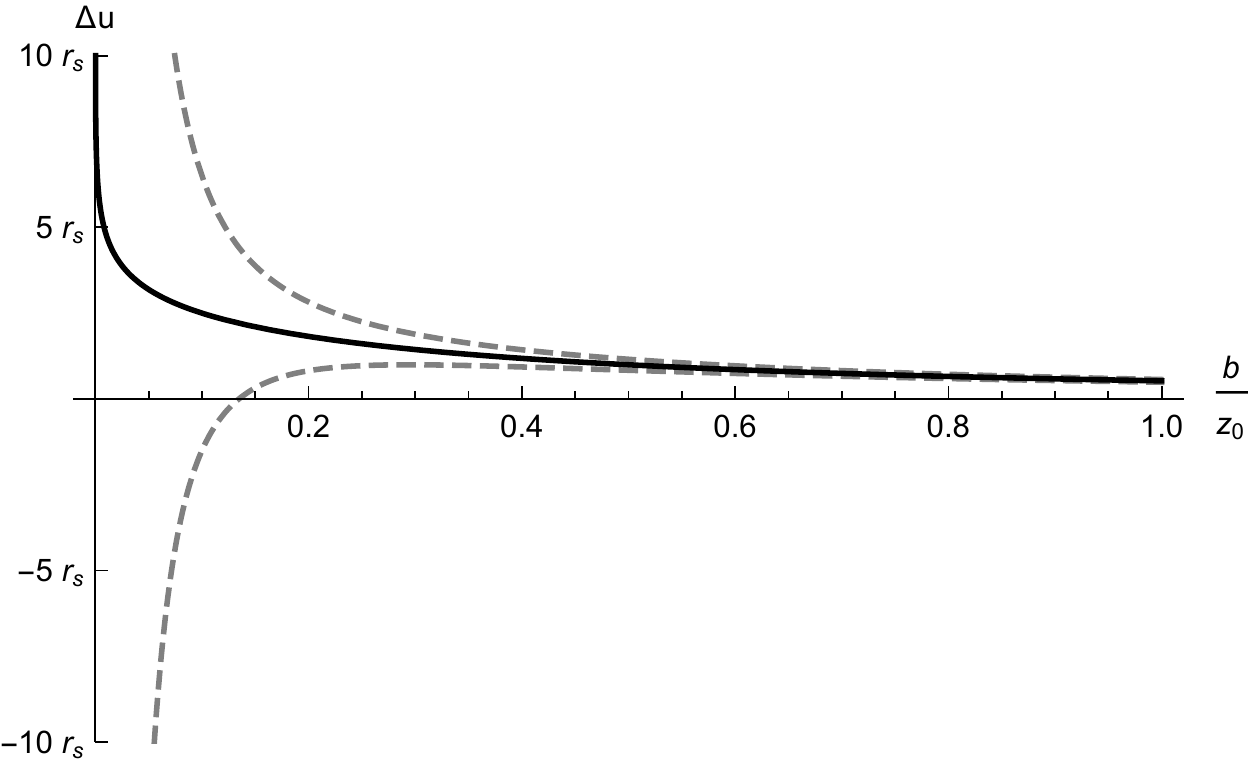,width=1\linewidth}
  \caption{Flat Spacetime}
  \label{fig:RFFflat}
\end{subfigure}
\begin{subfigure}{.5\textwidth}
  \centering
\epsfig{file=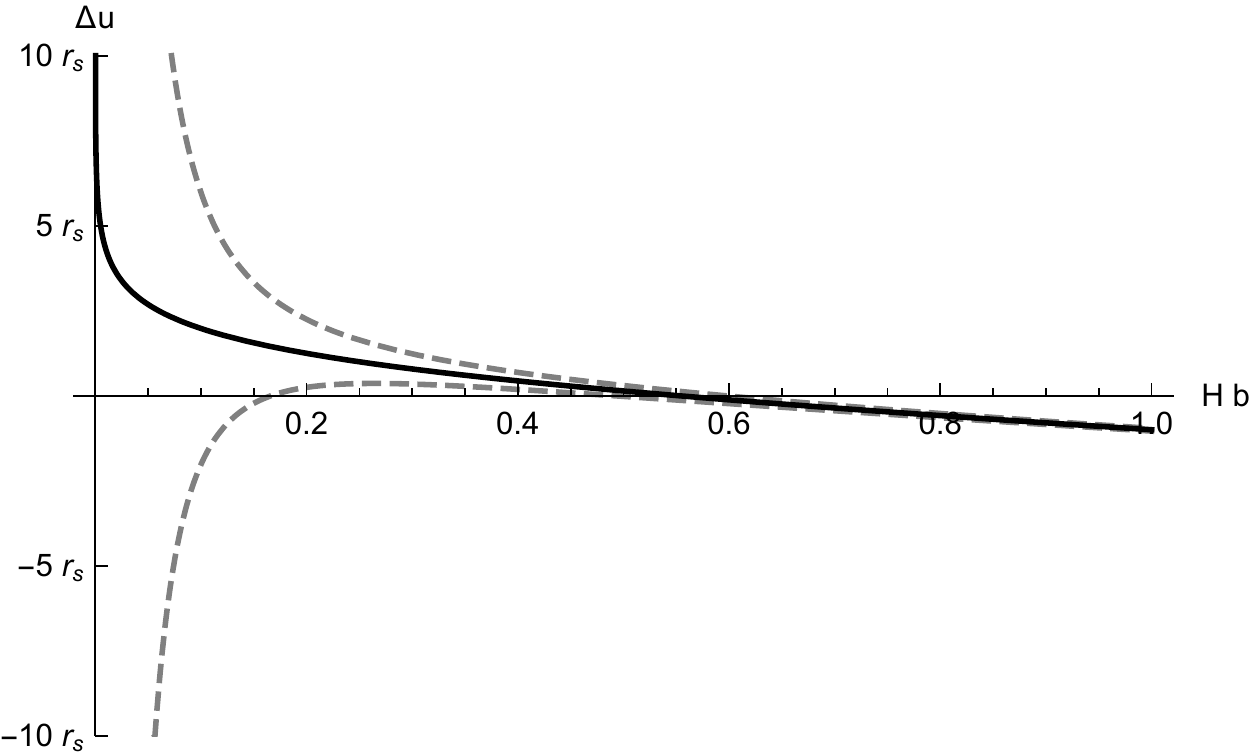,width=1\linewidth} 
  \caption{De Sitter Spacetime}
  \label{fig:RFFdS}
\end{subfigure}
\caption{Shifts $\Delta u$ for photons in a shockwave background as a function of impact parameter $b$. The black/solid line is the shift for the minimally coupling photon; the gray/dashed lines are the shifts for the two photon polarizations with an RFF coupling with $r_S\equiv 2Gp_u$. In de Sitter spacetime, the stretching of the Penrose diagram is responsible for $\Delta u$ being negative at large impact parameter $b$.  More problematic is the lowest dashed curve which is more negative at small $b$ than at large $b$.}
\end{figure}

\subsection{Non-Minimal Coupling}
By adding interactions, the light ray will no longer necessarily follow the null geodesic previously analyzed.  In particular, we now consider a non-minimally coupled massless spin-1 field\footnote{We are coupling $F_{\mu \nu}$ to the subtracted Riemann tensor, $\hat{R}_{\mu \nu\alpha\beta} \equiv R_{\mu \nu \alpha \beta} - R_{\mu \nu \alpha \beta}^{dS}$, which is equal to the Weyl tensor on-shell.  See \cite{Camanho:2014apa,Horowitz:1999gf} for more details. See also \cite{Camanho:2014apa} for the flat spacetime calculation.}
\begin{equation}
    S = \int d^{4}x\sqrt{-g}\Big(-\frac{1}{4}F^{\mu \nu}F_{\mu \nu} + \frac{\alpha}{4}\hat{R}_{\mu \nu}^{~~\alpha\beta}F^{\mu \nu}F_{\alpha\beta}\Big) \, ,
\end{equation}
with equations of motion
\begin{equation}
  \nabla^{\mu}(F_{\mu \nu}-\alpha \hat{R}_{\mu\nu}^{~~\alpha\beta}F_{\alpha\beta}) = 0 \, .
\end{equation}

The equations of motion for the two physical polarizations now differ from each other.  Using the same decomposition as above \eqref{decomp}, they are given by
\begin{align}
    &\partial_u\partial_vB(u,v)+\frac{H^2(1+k^2)}{(1 - H^2 uv)^2}B(u,v)  + 2Gp_u\,\delta(v)\,{\cal F}_B(b)\,\partial_{u}^2B(u,v) =0 \, ,\nonumber\\
    & \partial_u\partial_vC(u,v)+\frac{H^2(1+k^2)}{(1 - H^2 uv)^2}C(u,v) + 2Gp_u\,\delta(v)\,{\cal F}_C(b)\,\partial_{u}^2C(u,v) =0\, ,
\end{align}
where we have defined the functions ${\cal F}_{B,C}$ as follows:
\begin{align}
    &{\cal F}_B(b) =\frac{f(b)}{4(1 - H^2 b^2)}-\frac{\alpha}{4}\left(\frac{f'(b)}{b}+\frac{2H^2f(b)}{1-H^2b^2} \right) = \frac{f(b)}{4(1 -H^2 b^2)} - \frac{\alpha}{b^2\sqrt{1 - H^2 b^2}} \, ,\nonumber\\
    &{\cal F}_C(b) =\frac{f(b)}{4(1 - H^2 b^2)}+\frac{\alpha}{4}\left(\frac{f'(b)}{b}+\frac{2H^2f(b)}{1-H^2b^2} \right)  = \frac{f(b)}{4(1 - H^2b^2)} + \frac{\alpha}{b^2\sqrt{1 - H^2b^2}} \, .
\end{align}
We can again read off the phase shift in the $u$-coordinate.  Multiplying through once more by the rotation factor $\sqrt{1 - H^2 b^2}$ found in the previous sections, we have
\begin{equation}
    \Delta_{\pm} u \rightarrow  2G p_u \left(\frac{f(b)}{4\sqrt{1 - H^2b^2}} \pm \frac{\alpha}{b^2}\right) \, .
\end{equation}
These phase shifts are plotted in Figure \ref{fig:RFFdS} as a function of $b$.  Taking $H\rightarrow 0$, we recover the flat spacetime result which we plot in Figure \ref{fig:RFFflat} for comparison.  We note that the $\alpha$-dependent correction is identical in flat and de Sitter spacetimes.

The $\alpha$-dependent corrections come with opposite signs for the two polarizations, indicating that, no matter the sign of $\alpha$, one polarization will always experience a positive shift at small impact parameter while the other polarization will always experience a negative shift.  In particular, at small enough impact parameter, the negative shift will be more negative than its value at the maximum impact parameter $Hb=1$. Geodesics at small impact parameter then lie {\it outside} the light cone set by the geodesic at maximum impact parameter.  Thus, as in flat spacetime, we conclude that this coupling is pathological, unless new physics appears at the appropriate short distance.

Moreover, unlike the negative shift seen at large impact parameter, the negative shift at small impact parameter can be made arbitrarily large.  This means that the entire future boundary of de Sitter can be made causally accessible with this coupling, again suggesting a pathology.

\section{Shockwaves and the ANEC}
Finally, we comment briefly on the relation of the results obtained here to the average null energy condition (ANEC).  The form of the shockwave metric makes this comparison straightforward.  In particular, we write the shockwave metric as
\be
g_{\mu\nu} = \bar{g}_{\mu\nu}+F[x]k_\mu k_\nu \, ,
\ee
where $\bar{g}_{\mu\nu}$ is the background de Sitter spacetime.  Due to the Kerr-Schild form, the full, exact equations of motion are in fact linear in the ``perturbation:"
\be
\hat{{\cal O}} h_{\mu \nu} = -16 \pi G\, T_{\mu\nu} \, ,
\ee
where $h_{\mu\nu} = F[x]k_\mu k_\nu$.  

Let us again use the lightcone coordinates $(u,v,b,\phi)$ as in \eqref{pen} and let ${\cal G}$ be the Green's function such that
\be
\hat{{\cal O}}\, {\cal G}(b-b') = \delta(b-b')
\ee
The phase shift experienced by a null ray crossing the shock can be written as
\be
\Delta u = \frac{1}{4(1-H^2 b^2)} \int_{-\infty}^{+\infty}dv\,h_{vv} = -\frac{4 \pi G}{1-H^2 b^2}\int db'\, {\cal G}(b-b') \int_{-\infty}^{+\infty}dv\,T_{vv}  \, .
\ee
The term all the way on the right of the above expression is precisely the averaged null energy along a complete geodesic.  The Green's function is the  expression found previously in equation \eqref{Fb}, up to an overall negative constant ${\cal G}(b)\propto - f[b]$.  In the analogous calculation in flat spacetime, the Green's function is sign-definite and thus the sign of the shift is directly linked to the sign of the average null energy integrated along the geodesic.  Moreover, in flat spacetime the sign of the shift has a straightforward interpretation in terms of causality.  In de Sitter however, because \eqref{Fb} takes on negative values, we see that the shift can be negative while the averaged null energy remains positive and, as we have argued, the relation between the sign of the shift and the same causality constraint is not as direct.

\section{Discussion}
An intuitive notion of causality relates the angle of light cones in a perturbed spacetime to that of the vacuum spacetime.  Namely, in a well-behaved theory, we expect perturbations in the bulk of the spacetime to slow down nearby null rays and never speed them up.  To make this notion well-defined requires the use of asymptotic observables.  In this work we have argued that this notion can be extended to de Sitter spacetime despite obvious challenges. In particular, by comparing spatial ``shifts" on the boundary of de Sitter spacetime, we suggest that the relevant criteria is that the null geodesic at largest impact parameter in de Sitter should always be the fastest.  This can be defined using boundary observables in a coordinate invariant way and is a straighforward generalization of the causality criteria of flat and anti de Sitter spacetimes.   It remains to be seen whether such a criteria can be derived from more fundamental principles as it can in anti-de Sitter spacetime \cite{Gao:2000ga}.  We note that the considerations of this paper are purely classical; it would be interesting to see how they are impacted by quantum effects.  Moreover it would be interesting to relate these considerations to the ideas put forth in \cite{Aalsma:2020aib,Aalsma:2021kle}.  

\vskip.5cm

\bigskip
{\bf Acknowledgements}: We would like to thank Dionysios Anninos and Frederik Denef for interesting discussions and especially Austin Joyce for early collaboration.  This work was supported by DOE grant DE-SC0011941 and Simons Foundation Award Number 555117.

\bigskip 

\appendix

\section{Harmonics}
\label{SVT}
In this section, we will detail some of the properties of scalar and vector harmonics on the transverse space.  Our starting point is the de Sitter metric in the cylindrical light cone coordinates:
\be
ds^2 =-4\left(1-H^2 b^2 \right) \frac{du\,dv}{\left(1-H^2 u v \right)^2}+\frac{db^2}{1-H^2 b^2}+b^2 d\phi^2 \, .
\ee
Pulling out an overall factor, we can write this as
\be
ds^2 =4 \left(1-H^2 b^2 \right)\left[-\frac{du\,dv}{\left(1-H^2 u v \right)^2}+ dA^2\right]\, ,
\ee
where we have defined
\be
dA^2 =\frac{1}{4} \left(\frac{db^2}{(1-H^2 b^2)^2}+\frac{b^2}{1-H^2 b^2} d\phi^2\right) \, .
\ee
This is the metric of the hyperbolic plane.  This can be made apparent by performing the coordinate transformation
\be
b=\frac{2R}{1+H^2R^2} \, ,
\ee
which gives
\be
dA^2 =\frac{1}{(1-H^2R^2)^2} \left(dR^2+ R^2 d\phi^2\right) \, .
\ee

The scalar Laplacian is given by
\begin{equation}
    D^2 \equiv 4 (1 -H^2 b^2)\left((1 - H^2b^2)\partial_b^2 + \frac{1 - 2H^2b^2}{b}\partial_b  + \frac{1}{b^2}\partial_{\phi}^2 \right) \, .
    \label{AdSLap}
\end{equation}

\bigskip

\noindent \textbf{Scalar Harmonics: } We now consider the scalar harmonics $Y_{km}(b, \phi)$ of the $2d$ scalar Laplacian \eqref{AdSLap}.  The Laplacian is separable and the harmonics take the form
\be
Y_{km}(b, \phi) = y_{km}(b)e^{im\phi} \, ,
\ee
\be
y_{km}(b) =  C_1b^{|m|}(1 - H^2 b^2)^{\frac{1}{4}(1 + ik)}\,_{2}F_{1}\Big(\frac{1 + 2|m| +ik}{4}, \frac{3 + 2|m| +ik}{4}, 1+|m|, H^2b^2\Big) \, .
\ee
They obey the eigenfunction equation:
\begin{equation}
    D^2Y_{km} = -H^2(1 + k^2) Y_{km} \, .
\end{equation}

\bigskip

\noindent \textbf{Vector Harmonics: }There are two classes of vector harmonics, which may be distinguished by their parity. We refer to these two sectors as ``even" and ``odd". The even sector harmonics are
\begin{equation}
    Y_{I}(b, \phi) = D_{I}Y(b,\phi), \hspace{1 cm}I = \{b, \phi\},
\end{equation}
and they obey
\begin{equation}
    D^2 Y_{I} = -H^2(5+k^2)Y_{I}, \hspace{1 cm}D^{I}Y_{I} = -H^2(1 + k^2)Y \, .
\end{equation}
The odd sector harmonics are defined using the Levi-Civita tensor density:
\begin{equation}
    \epsilon_{IJ} = \frac{b}{4(1 - H^2 b^2)^{3/2}}\tilde{\epsilon}_{IJ},
\end{equation}
where $\tilde{\epsilon}_{IJ}$ is the antisymmetric Levi-Civita symbol. The odd harmonics are given by
\be
V_{I} = \epsilon_I ^{~J} D_{J}Y \, ,
\ee
so that
\begin{align}
    &V_{b} = \frac{1}{b(1 - H^2 b^2)^{1/2}}\partial_{\phi}Y, \hspace{1 cm}V_{\phi} = -b(1 - H^2 b^2)^{1/2}\partial_{b}Y \, .
\end{align}
They obey the following eigenfunction equation and tranversality constraint:
\begin{align}
    D^2 V_{I} = -H^2(5+k^2)V_{I}, \hspace{1 cm} D^{I}V_{I} = 0 \, .
\end{align}

\bibliographystyle{utphys}
\addcontentsline{toc}{section}{References}
\bibliography{dS_shapiro}

\providecommand{\href}[2]{#2}\begingroup\raggedright\begin{thebibliography}{10}

\bibitem{Camanho:2014apa}
X.~O. Camanho, J.~D. Edelstein, J.~Maldacena, and A.~Zhiboedov, ``{Causality
  Constraints on Corrections to the Graviton Three-Point Coupling},''
  \href{http://dx.doi.org/10.1007/JHEP02(2016)020}{{\em JHEP} {\bf 02} (2016)
  020}, \href{http://arxiv.org/abs/1407.5597}{{\tt arXiv:1407.5597 [hep-th]}}.

\bibitem{tHooft:1987vrq}
G.~'t~Hooft, ``{Graviton Dominance in Ultrahigh-Energy Scattering},''
  \href{http://dx.doi.org/10.1016/0370-2693(87)90159-6}{{\em Phys. Lett. B}
  {\bf 198} (1987)  61--63}.

\bibitem{Dray:1984ha}
T.~Dray and G.~'t~Hooft, ``{The Gravitational Shock Wave of a Massless
  Particle},'' \href{http://dx.doi.org/10.1016/0550-3213(85)90525-5}{{\em Nucl.
  Phys. B} {\bf 253} (1985)  173--188}.

\bibitem{Kabat:1992tb}
D.~N. Kabat and M.~Ortiz, ``{Eikonal quantum gravity and Planckian
  scattering},'' \href{http://dx.doi.org/10.1016/0550-3213(92)90627-N}{{\em
  Nucl. Phys. B} {\bf 388} (1992)  570--592},
  \href{http://arxiv.org/abs/hep-th/9203082}{{\tt arXiv:hep-th/9203082}}.

\bibitem{Visser:1998ua}
M.~Visser, B.~Bassett, and S.~Liberati, ``{Superluminal censorship},''
  \href{http://dx.doi.org/10.1016/S0920-5632(00)00782-9}{{\em Nucl. Phys. B
  Proc. Suppl.} {\bf 88} (2000)  267--270},
  \href{http://arxiv.org/abs/gr-qc/9810026}{{\tt arXiv:gr-qc/9810026}}.

\bibitem{Gao:2000ga}
S.~Gao and R.~M. Wald, ``{Theorems on gravitational time delay and related
  issues},'' \href{http://dx.doi.org/10.1088/0264-9381/17/24/305}{{\em Class.
  Quant. Grav.} {\bf 17} (2000)  4999--5008},
  \href{http://arxiv.org/abs/gr-qc/0007021}{{\tt arXiv:gr-qc/0007021}}.

\bibitem{Morris:1988tu}
M.~S. Morris, K.~S. Thorne, and U.~Yurtsever, ``{Wormholes, Time Machines, and
  the Weak Energy Condition},''
  \href{http://dx.doi.org/10.1103/PhysRevLett.61.1446}{{\em Phys. Rev. Lett.}
  {\bf 61} (1988)  1446--1449}.

\bibitem{Dubovsky:2005xd}
S.~Dubovsky, T.~Gregoire, A.~Nicolis, and R.~Rattazzi, ``{Null energy condition
  and superluminal propagation},''
  \href{http://dx.doi.org/10.1088/1126-6708/2006/03/025}{{\em JHEP} {\bf 03}
  (2006)  025}, \href{http://arxiv.org/abs/hep-th/0512260}{{\tt
  arXiv:hep-th/0512260}}.

\bibitem{Kelly:2014mra}
W.~R. Kelly and A.~C. Wall, ``{Holographic proof of the averaged null energy
  condition},'' \href{http://dx.doi.org/10.1103/PhysRevD.90.106003}{{\em Phys.
  Rev. D} {\bf 90} (2014) no.~10, 106003},
  \href{http://arxiv.org/abs/1408.3566}{{\tt arXiv:1408.3566 [gr-qc]}}.
  [Erratum: Phys.Rev.D 91, 069902 (2015)].

\bibitem{Engelhardt:2016aoo}
N.~Engelhardt and S.~Fischetti, ``{The Gravity Dual of Boundary Causality},''
  \href{http://dx.doi.org/10.1088/0264-9381/33/17/175004}{{\em Class. Quant.
  Grav.} {\bf 33} (2016) no.~17, 175004},
  \href{http://arxiv.org/abs/1604.03944}{{\tt arXiv:1604.03944 [hep-th]}}.

\bibitem{Hartman:2016lgu}
T.~Hartman, S.~Kundu, and A.~Tajdini, ``{Averaged Null Energy Condition from
  Causality},'' \href{http://dx.doi.org/10.1007/JHEP07(2017)066}{{\em JHEP}
  {\bf 07} (2017)  066}, \href{http://arxiv.org/abs/1610.05308}{{\tt
  arXiv:1610.05308 [hep-th]}}.

\bibitem{Aichelburg:1970dh}
P.~C. Aichelburg and R.~U. Sexl, ``{On the Gravitational field of a massless
  particle},'' \href{http://dx.doi.org/10.1007/BF00758149}{{\em Gen. Rel.
  Grav.} {\bf 2} (1971)  303--312}.

\bibitem{Hotta:1992qy}
M.~Hotta and M.~Tanaka, ``{Shock wave geometry with nonvanishing cosmological
  constant},'' \href{http://dx.doi.org/10.1088/0264-9381/10/2/012}{{\em Class.
  Quant. Grav.} {\bf 10} (1993)  307--314}.

\bibitem{Horowitz:1999gf}
G.~T. Horowitz and N.~Itzhaki, ``{Black holes, shock waves, and causality in
  the AdS / CFT correspondence},''
  \href{http://dx.doi.org/10.1088/1126-6708/1999/02/010}{{\em JHEP} {\bf 02}
  (1999)  010}, \href{http://arxiv.org/abs/hep-th/9901012}{{\tt
  arXiv:hep-th/9901012}}.

\bibitem{Aalsma:2020aib}
L.~Aalsma and G.~Shiu, ``{Chaos and complementarity in de Sitter space},''
  \href{http://dx.doi.org/10.1007/JHEP05(2020)152}{{\em JHEP} {\bf 05} (2020)
  152}, \href{http://arxiv.org/abs/2002.01326}{{\tt arXiv:2002.01326
  [hep-th]}}.

\bibitem{Aalsma:2021kle}
L.~Aalsma, A.~Cole, E.~Morvan, J.~P. van~der Schaar, and G.~Shiu, ``{Shocks and
  information exchange in de Sitter space},''
  \href{http://dx.doi.org/10.1007/JHEP10(2021)104}{{\em JHEP} {\bf 10} (2021)
  104}, \href{http://arxiv.org/abs/2105.12737}{{\tt arXiv:2105.12737
  [hep-th]}}.

\end{thebibliography}\endgroup

\end{document}